\documentclass[11pt,a4paper]{article}
\usepackage[utf8]{inputenc}
\usepackage[T1]{fontenc}
\usepackage{amsmath,amssymb,amsfonts}
\usepackage{graphicx}
\usepackage{booktabs}
\usepackage{multirow}
\usepackage{url}
\usepackage{hyperref}
\usepackage[margin=1in]{geometry}
\usepackage{caption}
\usepackage{subcaption}
\usepackage{enumitem}
\usepackage{xcolor}
\usepackage{listings}
\usepackage{float}

\title{A Unified Compliance Aggregator Framework for
Automated Multi-Tool Security Assessment of Linux Systems}

\author{
\textbf{Sheldon Paul}\\
Department of Computing, Engineering and Mathematical
Science\\
Texas A\&M University - San Antonio, TX, USA\\
\texttt{sthil01@jaguar.tamu.edu}
\and
\textbf{Izzat Alsmadi}\\
Professor, Department of Computing, Engineering and
Mathematical Science\\
Texas A\&M University - San Antonio, TX, USA\\
\texttt{ialsmadi@tamusa.edu}
}

\date{}

\begin{document}

\maketitle

\begin{abstract}
Assessing the security posture of modern computing systems typically requires the use of multiple specialized tools. These tools focus on different aspects such as configuration compliance, file integrity, and vulnerability exposure, and their outputs are often difficult to interpret collectively. This paper introduces the Unified Compliance Aggregator (UCA), a framework that integrates several open-source security tools into a single composite score representing overall system security.

The proposed framework combines outputs from Lynis, OpenSCAP (STIG and CIS profiles), AIDE, Tripwire, and Nmap NSE. A normalization process converts heterogeneous outputs into a consistent 0–100 scale, followed by weighted aggregation. We also introduce a logarithmic scoring model for file integrity measurements to address limitations observed in prior linear approaches.

Experiments were conducted on Ubuntu 22.04 across different hardening levels and environments. Results show consistent improvement in composite scores as systems are hardened, while also revealing contrasting behavior between compliance and file integrity tools. Two case studies—a basic web server and a DVWA-based system—illustrate how the framework can be applied in practical scenarios.

\end{abstract}

\textbf{Keywords:} security auditing, compliance
aggregation, system hardening, Linux security, CIS
benchmarks, file integrity monitoring, vulnerability
scanning, FABRIC testbed, DVWA

\section{Introduction}

Security assessments are typically performed using multiple independent tools, each targeting a specific domain such as system configuration, vulnerability scanning, or file integrity monitoring. While each tool provides useful information, their outputs are fragmented and often difficult to interpret together.

For example, a system administrator may rely on Lynis for configuration auditing, OpenSCAP for compliance checking, and AIDE or Tripwire for file integrity monitoring. Each tool produces its own metrics, formats, and scales, making it challenging to form a unified understanding of the system’s overall security posture.

This fragmentation motivates the need for a unified framework that can integrate outputs from multiple tools and provide a consolidated, interpretable measure of security.

Standards such as PCI DSS, HIPAA, FedRAMP, and SOC~2 impose
requirements spanning configuration management, access
control, and vulnerability management simultaneously. No
single tool covers all domains. The research objective is to
develop a framework that aggregates outputs from multiple
security tools, normalizes results to a common scale, and
produces a single actionable score reflecting the overall
security posture of particularly for Linux operating systems.

\subsection{Research Questions}

\begin{enumerate}[label=\textbf{RQ\arabic*:}]
\item How can multiple security tools with different output
formats be integrated into a unified framework?
\item What normalization methodologies convert diverse outputs
into comparable scores?
\item How should weighted scoring account for differences in
tools scope?
\item How does progressive hardening affect the composite
score across multiple dimensions?
\item What is the practical impact when applied to web server
and web application server deployments?
\end{enumerate}

\subsection{Contributions}

Our previous work \cite{paul2025arxiv} introduced UCA using
three tools (Lynis, OpenSCAP, AIDE) on FABRIC (\cite{fabric2019}), achieving
statistically significant results across 108 audit runs.
Three limitations remained: no vulnerability assessment,
single platform only, and a linear AIDE formula that
collapsed beyond 20 file changes.

This paper contributes: (1)~extended tool integration from
three to six; (2)~firewall activation identified as the
single largest improvement at +47 points; (3)~cross-platform
validation across VMware and FABRIC; (4)~logarithmic AIDE
scoring preventing collapse; (5)~a basic web server case
study (+8.4\%); (6)~a DVWA web application server case study
(+3.8\%) using a recognized intentionally vulnerable
platform; and (7)~empirical evidence that compliance and file
integrity tools exhibit opposite trends during hardening.

\section{Related Work}

\subsection{Security Auditing Tools}

Lynis \cite{lynis} performs comprehensive system audits
across 300+ tests producing a Hardening Index (0-100).
OpenSCAP \cite{openscap} implements SCAP \cite{scap},
supporting DISA STIG and CIS profiles through the SCAP
Security Guide \cite{ssg}. CIS Benchmarks \cite{cis} provide
industry-consensus configurations referenced by PCI DSS,
HIPAA, and FedRAMP. AIDE \cite{aide} and Tripwire
\cite{tripwire} provide file integrity monitoring mandated by
PCI DSS Requirement~11.5 and NIST SP~800-53 SI-7
\cite{nist800-53}. Nmap \cite{nmap} with NSE scripts provides
network vulnerability detection; OpenVAS \cite{openvas}
offers CVSS-based enterprise scanning.

\subsection{Web Application Security and DVWA}

Web applications introduce risks cataloged by the OWASP
Top~10 \cite{owasp}. DVWA (Damn Vulnerable Web Application)
\cite{dvwa}, endorsed by OWASP with 8,000+ GitHub stars,
provides a controlled environment with documented SQL
injection, XSS, command injection, file inclusion, file
upload, CSRF, and brute force vulnerabilities. Used in CEH
and OSCP certification courses and university curricula
worldwide, it is a credible reproducible research target.

\subsection{Integration Approaches}

SIEM systems \cite{siem} aggregate events for monitoring but
do not produce compliance scores. SOAR platforms \cite{soar}
automate incident response without composite metrics.
Enterprise platforms (Tenable.io, Qualys VMDR) offer
integration but are proprietary \cite{commercial_comparison}.
Academic work has explored multi-criteria assessment
\cite{multicriteria} and automated compliance checking
\cite{autocomp}. While their are tools and research publications that focus on the integration of several tools or frameworks to combine the strength of all those tools, we believe that their is still a great value of evaluating the integration of several security tools and evaluate the overall value of such integration. Limited open-source, reproducible frameworks that unify heterogeneous tool outputs into a normalized composite score with empirical validation.

\section{Methodology}

\subsection{Framework Architecture}

\begin{figure}[h]
\centering
\includegraphics[width=\textwidth]
{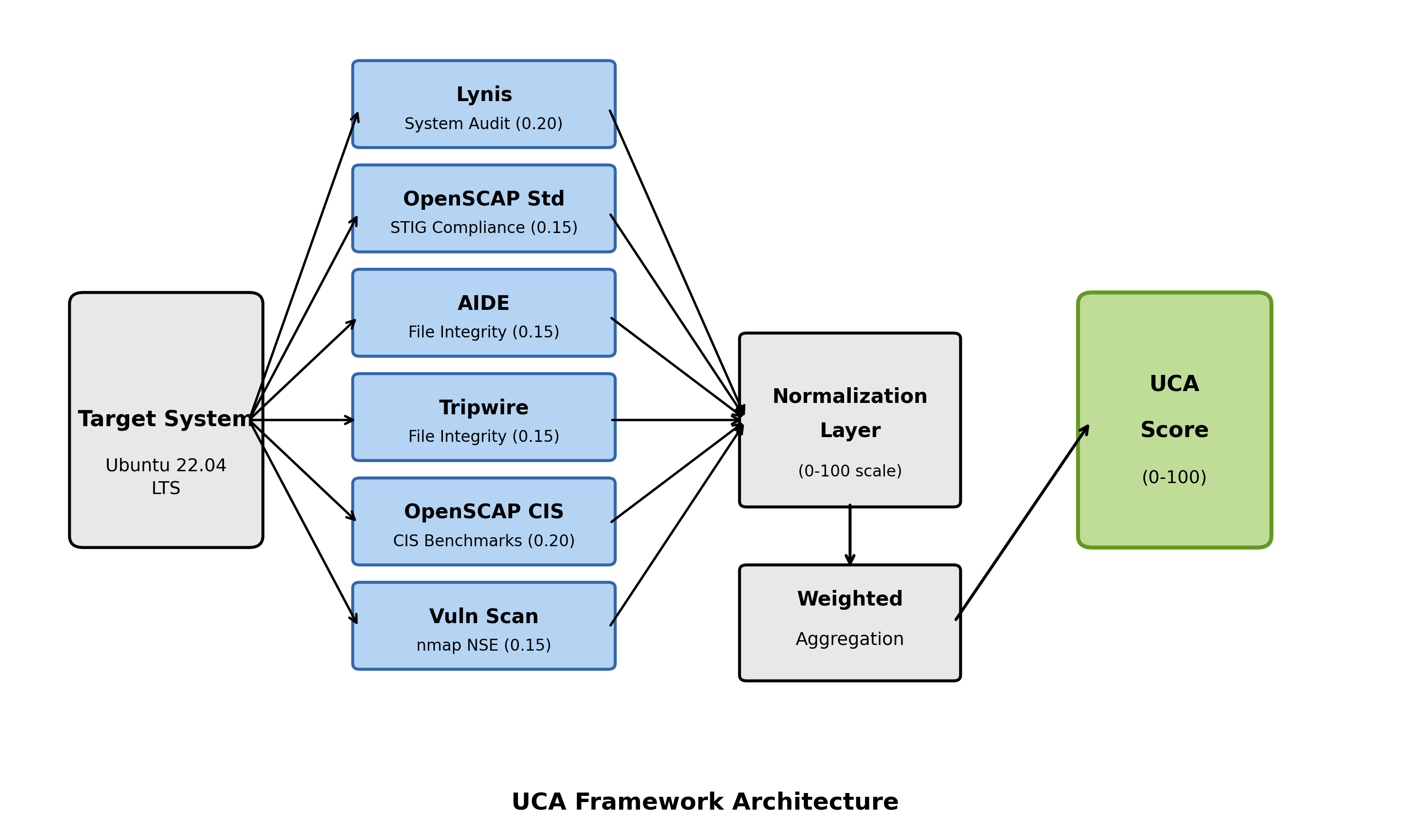}
\caption{UCA Framework Architecture. Six tools independently
scan the target, outputs are normalized to 0-100, and
aggregated through weighted summation into a composite UCA
score.}
\label{fig:arch}
\end{figure}

The UCA framework operates through four phases:
(1)~tool deployment and database initialization before
hardening; (2)~scan execution through CLI interfaces;
(3)~output normalization to 0-100; and (4)~weighted
aggregation into a composite score. Figure~\ref{fig:arch}
illustrates the architecture.

Its important to acknowledge some facts and limitations about UCA. UCA is not a definitive measure of security but a comparative and operational metric.

\subsection{Tool Selection}

Six tools were selected based on open-source availability,
Ubuntu support, quantifiable output, complementary domain
coverage, and active maintenance (Table~\ref{tab:tools}).
Two file integrity tools provide redundant cross-validation;
two OpenSCAP profiles provide simultaneous STIG (43~rules)
and CIS Level~1 Server (247~rules) coverage.

\begin{table}[h]
\centering
\caption{UCA Tool Suite}
\label{tab:tools}
\begin{tabular}{llll}
\toprule
\textbf{Tool} & \textbf{Ver.} & \textbf{Domain}
& \textbf{Output} \\
\midrule
Lynis & 3.0.7 & System audit & Index (0-100) \\
OpenSCAP Std & 1.2.17 & STIG compliance &
Pass/Fail counts \\
AIDE & 0.17.4 & File integrity & Change counts \\
Tripwire & 2.4.3.7 & File integrity &
Object/Violation counts \\
OpenSCAP CIS & 1.2.17 & CIS benchmarks &
Pass/Fail counts \\
nmap NSE & 7.80 & Vulnerability &
CVE/Port findings \\
\bottomrule
\end{tabular}
\end{table}

\subsection{Score Normalization}

\textbf{Lynis:} $S_{Lynis} = HardeningIndex$, directly
extracted (0-100).

\textbf{OpenSCAP (Standard and CIS):}
\begin{equation}
S_{SCAP} = \frac{P_{pass}}{P_{pass} + P_{fail}} \times 100
\end{equation}

\textbf{AIDE Logarithmic Scoring:} The previous linear
formula ($S = 100 - 5V$) collapsed beyond 20 changes. The
logarithmic replacement maintains sensitivity across
hundreds of legitimate hardening-induced modifications:
\begin{equation}
S_{AIDE} = \max\!\left(0,\;
100 - 10 \cdot \log_{10}(V_{total})\right)
\end{equation}

\textbf{Tripwire:}
\begin{equation}
S_{TW} = \frac{O_{total} - V_{total}}{O_{total}} \times 100
\end{equation}

\textbf{Vulnerability Scan:}
\begin{equation}
S_{Vuln} = \max\!\left(0,\; 100 -
\textstyle\sum_{i} w_i n_i - P_{ports} - P_{vulns}\right)
\end{equation}
Severity weights: critical ($w{=}15$), high ($w{=}8$),
medium ($w{=}4$), low ($w{=}1$); open port penalty 3 per
port; confirmed vulnerability penalty 10 each; active
firewall reduces total penalty by 10.

\subsection{Weighted Aggregation}

\begin{equation}
UCA = \sum_{i=1}^{6} w_i \cdot S_i, \quad
\sum w_i = 1.00
\end{equation}

\begin{figure}[h]
\centering
\includegraphics[width=0.5\textwidth]
{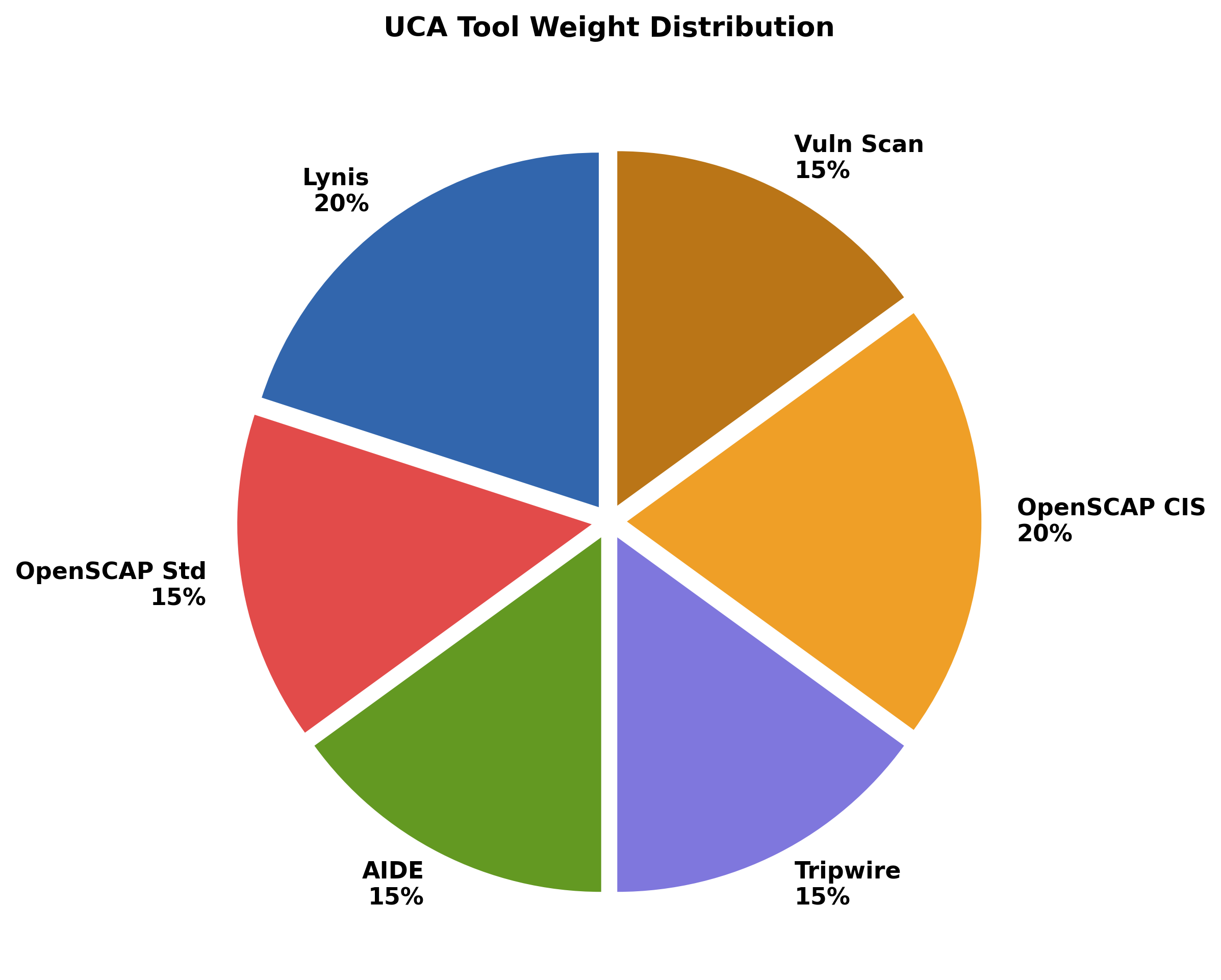}
\caption{UCA weight distribution. Multi-domain tools (Lynis,
OpenSCAP CIS) receive 0.20 each; single-domain tools receive
0.15, reflecting scope-based assignment.}
\label{fig:weights}
\end{figure}

Multi-domain tools receive weight 0.20; single-domain tools
receive 0.15 (Table~\ref{tab:weights},
Figure~\ref{fig:weights}). We acknowledge that such weights can be subjective and may require further investigations. As an alternative, they can be mapped to CVSS base scores.

\begin{table}[h]
\centering
\caption{UCA Weight Configuration}
\label{tab:weights}
\begin{tabular}{lcp{6cm}}
\toprule
\textbf{Tool} & \textbf{Weight} & \textbf{Rationale} \\
\midrule
Lynis & 0.20 & Multi-domain: 300+ tests across auth,
networking, kernel, services \\
OpenSCAP Standard & 0.15 & Single-focus: STIG (43 rules) \\
AIDE & 0.15 & Single-domain: file integrity \\
Tripwire & 0.15 & Single-domain: file integrity \\
OpenSCAP CIS & 0.20 & Multi-domain: CIS Level~1 Server
(247 rules) \\
Vulnerability & 0.15 & Single-domain: network exposure \\
\midrule
\textbf{Total} & \textbf{1.00} & \\
\bottomrule
\end{tabular}
\end{table}

\subsection{Hardening Specifications}

Three progressive hardening levels were defined
(Table~\ref{tab:hardening}).

\begin{table}[h]
\centering
\caption{Hardening Level Specifications}
\label{tab:hardening}
\begin{tabular}{lp{10cm}}
\toprule
\textbf{Level} & \textbf{Configuration} \\
\midrule
Baseline & Default Ubuntu 22.04 LTS, no modifications \\
\midrule
Partial & SSH: root login disabled, password auth disabled,
MaxAuthTries~3, ClientAlive 300s/2. UFW: deny incoming, allow
SSH. CUPS and Avahi disabled. Login banners on /etc/issue,
/etc/issue.net \\
\midrule
Full & All partial plus: sysctl hardening (IP forwarding off,
SYN cookies, ASLR max, SUID dump off, ICMP redirects blocked,
reverse path filtering, martian logging). auditd rules for
/etc/passwd, /etc/shadow, /etc/group, /etc/sudoers. USB
blacklisted. Password aging 90/7/14 days, UMASK~027. Core
dumps disabled. Filesystems blacklisted: cramfs, freevxfs,
jffs2, hfs, hfsplus, squashfs, udf. SSH: X11/TCP forwarding
off, LoginGraceTime~60s, MaxStartups 10:30:60. Banners on
/etc/issue, /etc/issue.net, /etc/motd \\
\bottomrule
\end{tabular}
\end{table}

\subsection{File Integrity Methodology}

AIDE and Tripwire databases are initialized once on the
unmodified baseline before any hardening and never
reinitialized. As hardening progresses, detected file changes
increase and integrity scores decrease, the consistent and
expected behavior. The weighted UCA formula accounts for this
inverse relationship in the composite score.

\section{Experimental Setup}

\subsection{FABRIC Testbed}

Experiments were conducted on the FABRIC research testbed
\cite{fabric2019}, a national-scale programmable
infrastructure. The FIU site was used, one of several
geographically distributed FABRIC infrastructure locations,
analogous to the EDUKY site used in our previous work
\cite{paul2025arxiv}. Five nodes were provisioned on a
private Layer~2 bridge network (10.10.1.0/24).

\begin{table}[h]
\centering
\caption{FABRIC Node Specifications}
\label{tab:nodes}
\begin{tabular}{lllp{3.5cm}}
\toprule
\textbf{Node} & \textbf{IP} & \textbf{Resources}
& \textbf{Role} \\
\midrule
Baseline & 10.10.1.1 & 4c/8GB/30GB & Unhardened control \\
Partial & 10.10.1.2 & 4c/8GB/30GB & SSH + Firewall \\
Full & 10.10.1.3 & 4c/8GB/30GB & Comprehensive \\
Scanner & 10.10.1.100 & 4c/8GB/40GB & Vuln scanner \\
DVWA & 10.10.1.4 & 4c/8GB/30GB & Web app case study \\
\bottomrule
\end{tabular}
\end{table}

All nodes run Ubuntu 22.04 LTS. The DVWA node was added to
the existing slice for Experiment~3. OpenVAS GVM~21.4 was
initially planned but exhibited persistent CLI socket errors;
nmap~7.80 with NSE scripts was adopted permanently.

\begin{figure}[h]
\centering
\includegraphics[width=\textwidth]{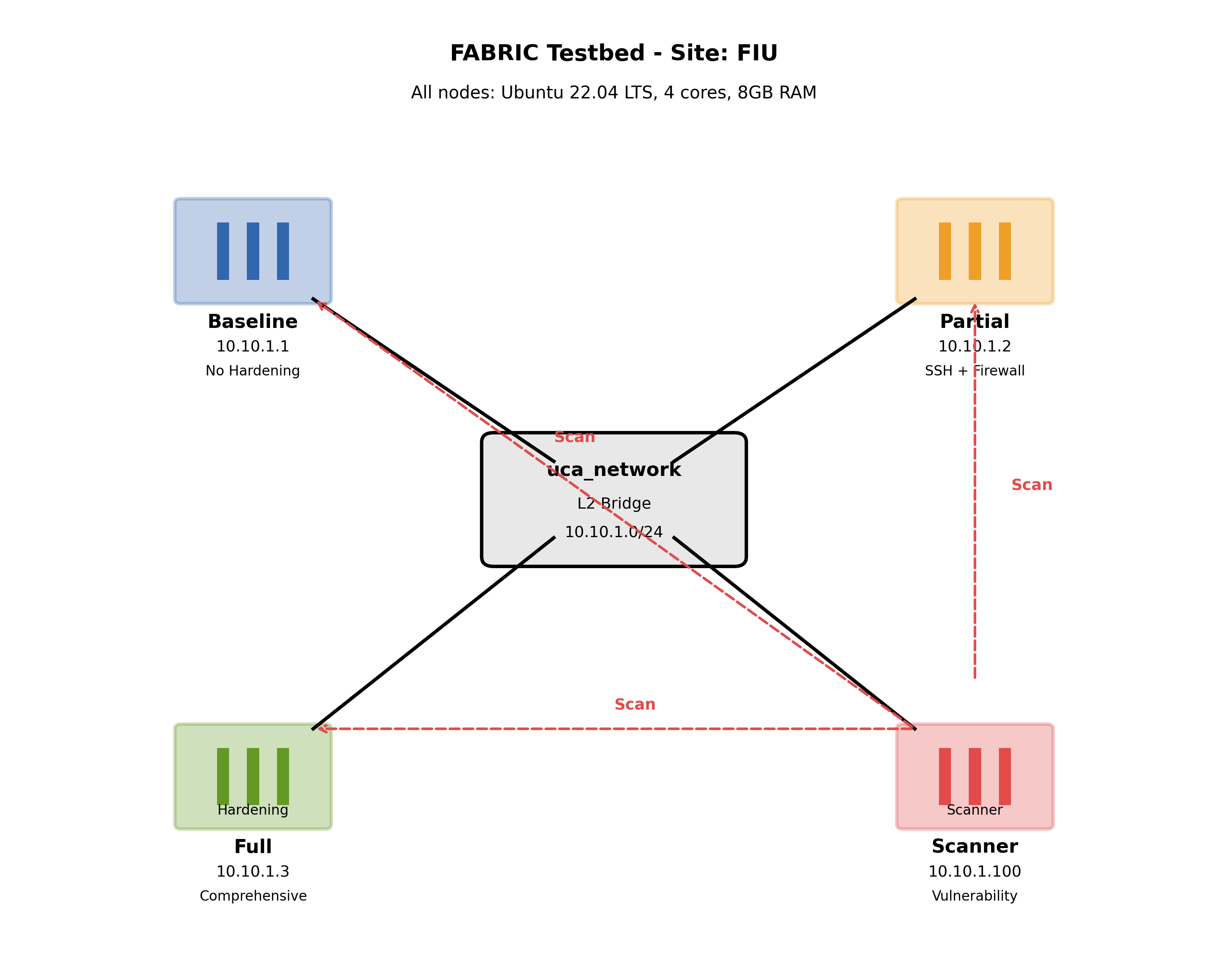}
\caption{FABRIC testbed topology. Five nodes on a private L2
bridge (10.10.1.0/24): three hardening targets, one scanner,
and one DVWA node for the web application case study.}
\label{fig:topology}
\end{figure}

\subsection{Local VM and Cross-Platform Configuration}

Cross-platform validation used VMware Workstation with
Ubuntu 22.04.5 LTS (2~cores, 4~GB RAM, 50~GB) using
snapshots at each hardening stage.

\begin{table}[h]
\centering
\caption{Cross-Platform Configuration Comparison}
\label{tab:crossplat}
\begin{tabular}{lll}
\toprule
\textbf{Parameter} & \textbf{Local VM} & \textbf{FABRIC} \\
\midrule
OS & Ubuntu 22.04.5 & Ubuntu 22.04 \\
CPU / RAM & 2 cores / 4 GB & 4 cores / 8 GB per node \\
Strategy & Snapshots & 5 independent VMs \\
Network & Localhost & L2 Bridge (10.10.1.0/24) \\
\bottomrule
\end{tabular}
\end{table}

AIDE scans required 30-45 minutes on the 4~GB local VM but
completed in approximately 5 minutes on the 8~GB FABRIC
nodes.

\subsection{Datasets and Resources}

Table~\ref{tab:datasets} summarizes external data
dependencies. SCAP content was ComplianceAsCode v0.1.72
(ssg-ubuntu2204-ds.xml, 14~MB SCAP~1.3 DataStream),
downloaded from GitHub as it was unavailable in Ubuntu
package repositories on FABRIC. Two profiles were used:
Standard (43~rules from DISA STIG) and CIS Level~1 Server
(247~rules). Vulnerability data utilized NVD through nmap
NSE scripts. Hardening configurations derived from CIS
Benchmarks, DISA STIGs, NIST SP~800-53, and OWASP
guidelines.

\begin{table}[h]
\centering
\caption{External Data Dependencies}
\label{tab:datasets}
\begin{tabular}{llp{4.5cm}}
\toprule
\textbf{Category} & \textbf{Source} & \textbf{Purpose} \\
\midrule
SCAP Content & ComplianceAsCode v0.1.72 &
OpenSCAP compliance evaluation \\
CIS Benchmarks & Center for Internet Security &
Industry security baselines \\
CVE Data & NVD / NIST & Vulnerability identification \\
OS & Ubuntu 22.04 LTS & Target system \\
Infrastructure & FABRIC Testbed \cite{fabric2019} &
Experimental platform \\
Standards & CIS, DISA, NIST, OWASP &
Hardening configuration guidance \\
\bottomrule
\end{tabular}
\end{table}

\section{Results and Analysis}

\subsection{Progressive Hardening Results}

Table~\ref{tab:results} presents complete six-tool results
across all hardening levels.

\begin{table}[h]
\centering
\caption{FABRIC Testbed:Complete Assessment Results}
\label{tab:results}
\begin{tabular}{lcccc}
\toprule
\textbf{Tool} & \textbf{Baseline} & \textbf{Partial}
& \textbf{Full} & \textbf{Change} \\
\midrule
Lynis & 59 & 61 & 66 & +11.9\% \\
OpenSCAP Standard & 67.4 & 69.8 & 77.3 & +14.7\% \\
AIDE & 83.4 & 77.7 & 75.0 & $-$8.4 pts \\
Tripwire & 82.4 & 78.0 & 77.7 & $-$4.7 pts \\
OpenSCAP CIS & 57.8 & 58.6 & 67.1 & +16.1\% \\
Vulnerability & 0 & 47 & 47 & +47 pts \\
\midrule
\textbf{UCA} & \textbf{58.34} & \textbf{64.80}
& \textbf{68.17} & \textbf{+16.8\%} \\
\bottomrule
\end{tabular}
\end{table}

\begin{figure}[h]
\centering
\includegraphics[width=\textwidth]{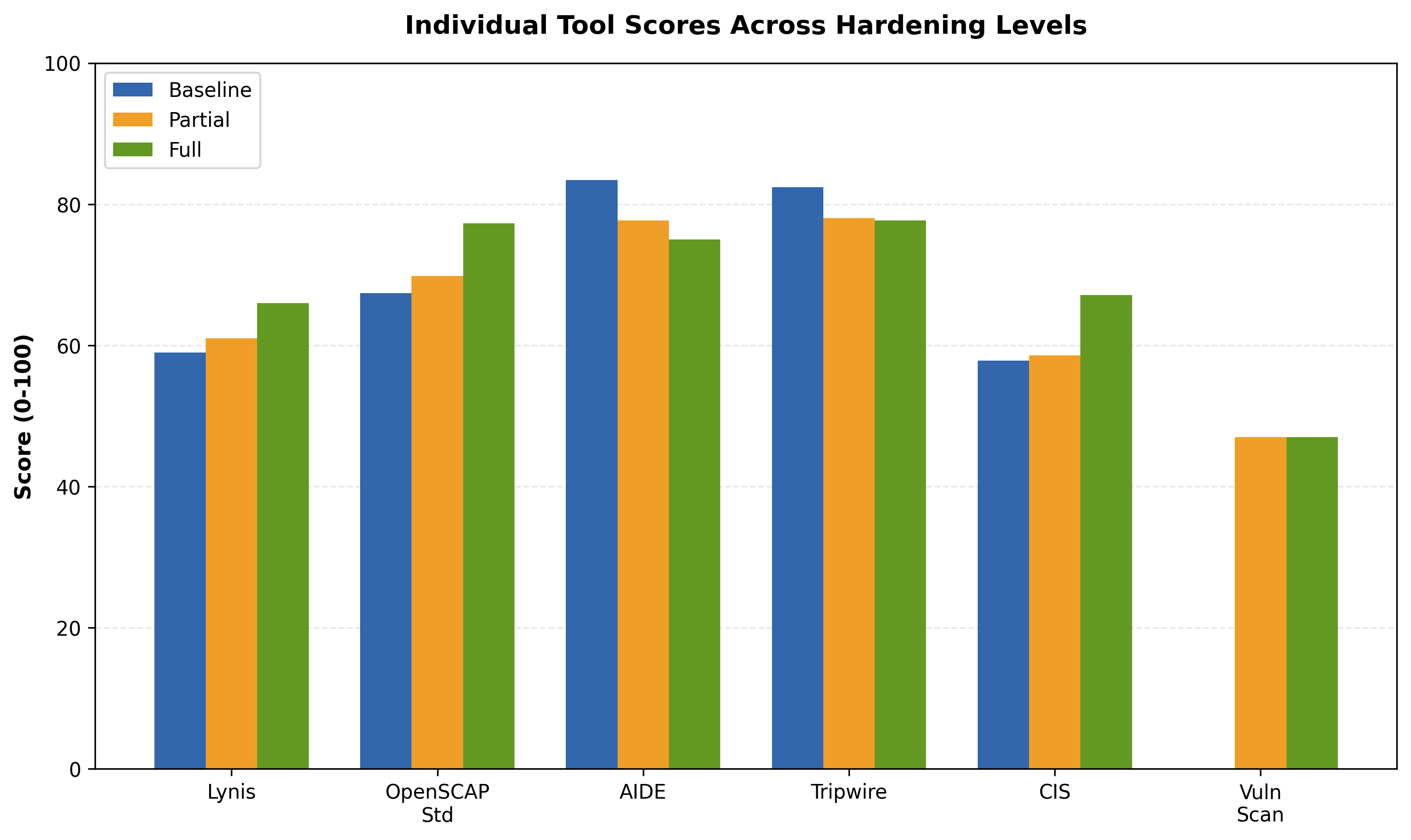}
\caption{Individual tool scores across three hardening
levels. Compliance tools trend upward; file integrity tools
trend downward reflecting legitimate change detection.}
\label{fig:scores}
\end{figure}

\subsubsection{Compliance Tools}
Lynis rose from 59 to 66 (+11.9\%) reflecting SSH, firewall,
kernel, and auditd hardening. OpenSCAP Standard improved from
67.4\% (29/43) to 77.3\% (34/43), +14.7\%. OpenSCAP CIS
showed the strongest gain from 57.8\% (137/237) to 67.1\%
(163/243), +16.1\%, with 26 additional rules satisfied
through kernel and filesystem restrictions.

\subsubsection{File Integrity Tools}

\begin{table}[h]
\centering
\caption{AIDE Detailed Change Detection}
\label{tab:aide}
\begin{tabular}{lccccc}
\toprule
\textbf{Level} & \textbf{Added} & \textbf{Removed}
& \textbf{Changed} & \textbf{Total} & \textbf{Score} \\
\midrule
Baseline & 11 & 0 & 35 & 46 & 83.4 \\
Partial & 6 & 0 & 165 & 171 & 77.7 \\
Full & 129 & 0 & 188 & 317 & 75.0 \\
\bottomrule
\end{tabular}
\end{table}

The logarithmic formula limited the AIDE score decrease to
8.4 points despite a sevenfold violation increase from
baseline to full hardening. Tripwire scanned 76,000+ objects
reporting 13,459 violations at baseline (82.4\%) increasing
to 17,540 after full hardening (77.7\%). Agreement between
both tools cross-validates the change detection findings.

\subsubsection{Vulnerability Assessment}

\begin{table}[h]
\centering
\caption{Vulnerability Scan Results}
\label{tab:vuln}
\begin{tabular}{lccc}
\toprule
\textbf{Metric} & \textbf{Baseline} & \textbf{Partial}
& \textbf{Full} \\
\midrule
Open Ports & 2 & 1 & 1 \\
Firewall Active & No & Yes & Yes \\
Filtered Ports & 0 & 65,534 & 65,534 \\
Confirmed Vulns & 4 & 0 & 0 \\
Vulnerability Score & 0 & 47 & 47 \\
\bottomrule
\end{tabular}
\end{table}

The baseline scored 0 with 2 open ports, no firewall, and 4
confirmed vulnerabilities. Partial hardening activated UFW,
filtering 65,534 ports and eliminating all confirmed
vulnerabilities. The score plateaued because remaining CVEs
(OpenSSH 8.9p1: CVE-2023-38408, CVE-2024-6387) require
software upgrades rather than configuration changes.

\subsection{UCA Composite Analysis}

\begin{figure}[h]
\centering
\includegraphics[width=0.8\textwidth]{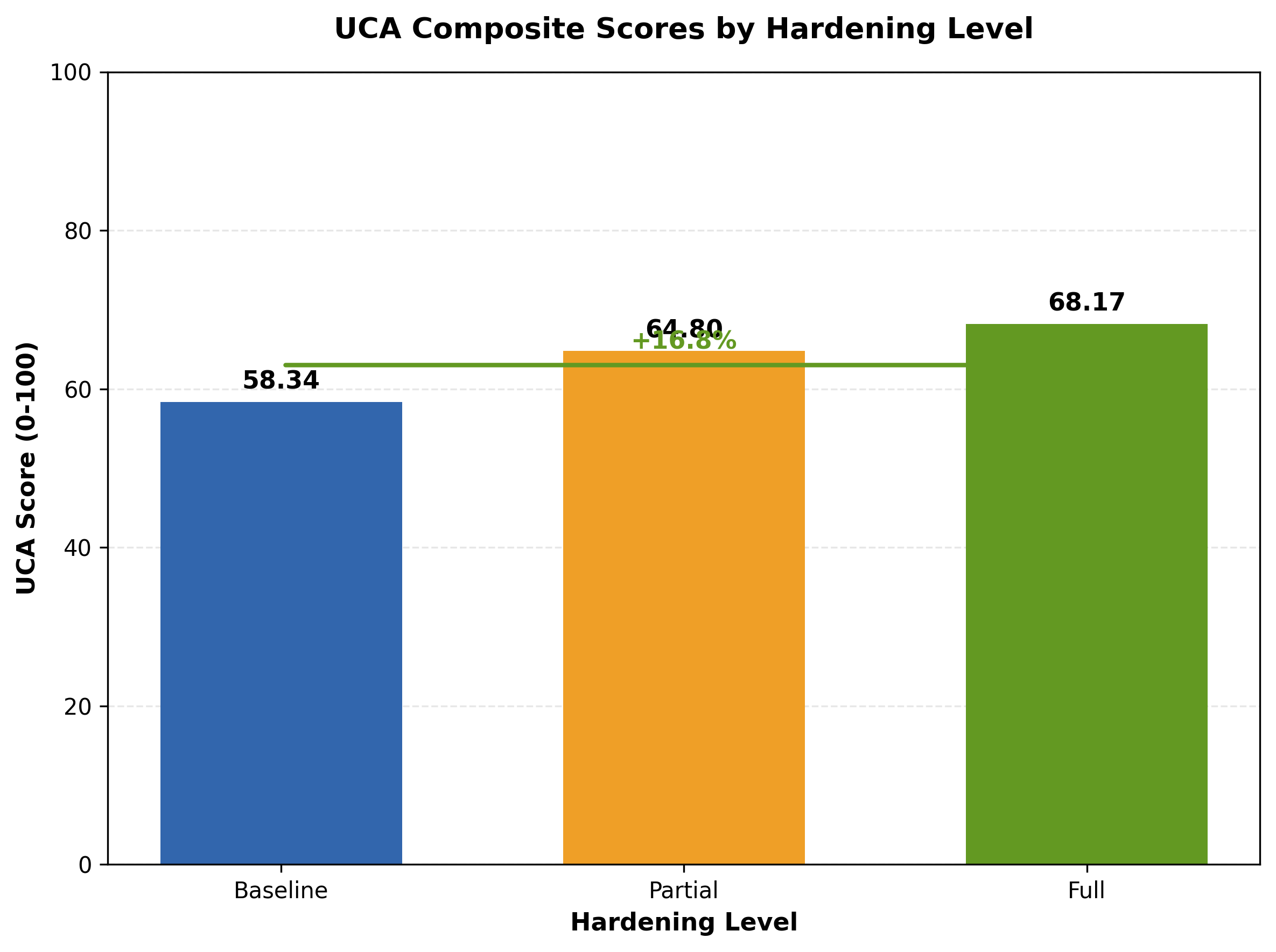}
\caption{UCA composite scores: 58.34 (baseline), 64.80
(partial), 68.17 (full), representing +16.8\% total
improvement.}
\label{fig:uca}
\end{figure}

The +9.83 point total improvement decomposes as: vulnerability
+7.05 pts (71.7\%), CIS +1.86 pts, Standard +1.49 pts, Lynis
+1.40 pts, Tripwire $-$0.71 pts, AIDE $-$1.26 pts. Firewall
activation alone accounted for nearly three-quarters of the
total improvement.

\subsection{Multi-Dimensional Behavior}

\begin{figure}[h]
\centering
\includegraphics[width=\textwidth]{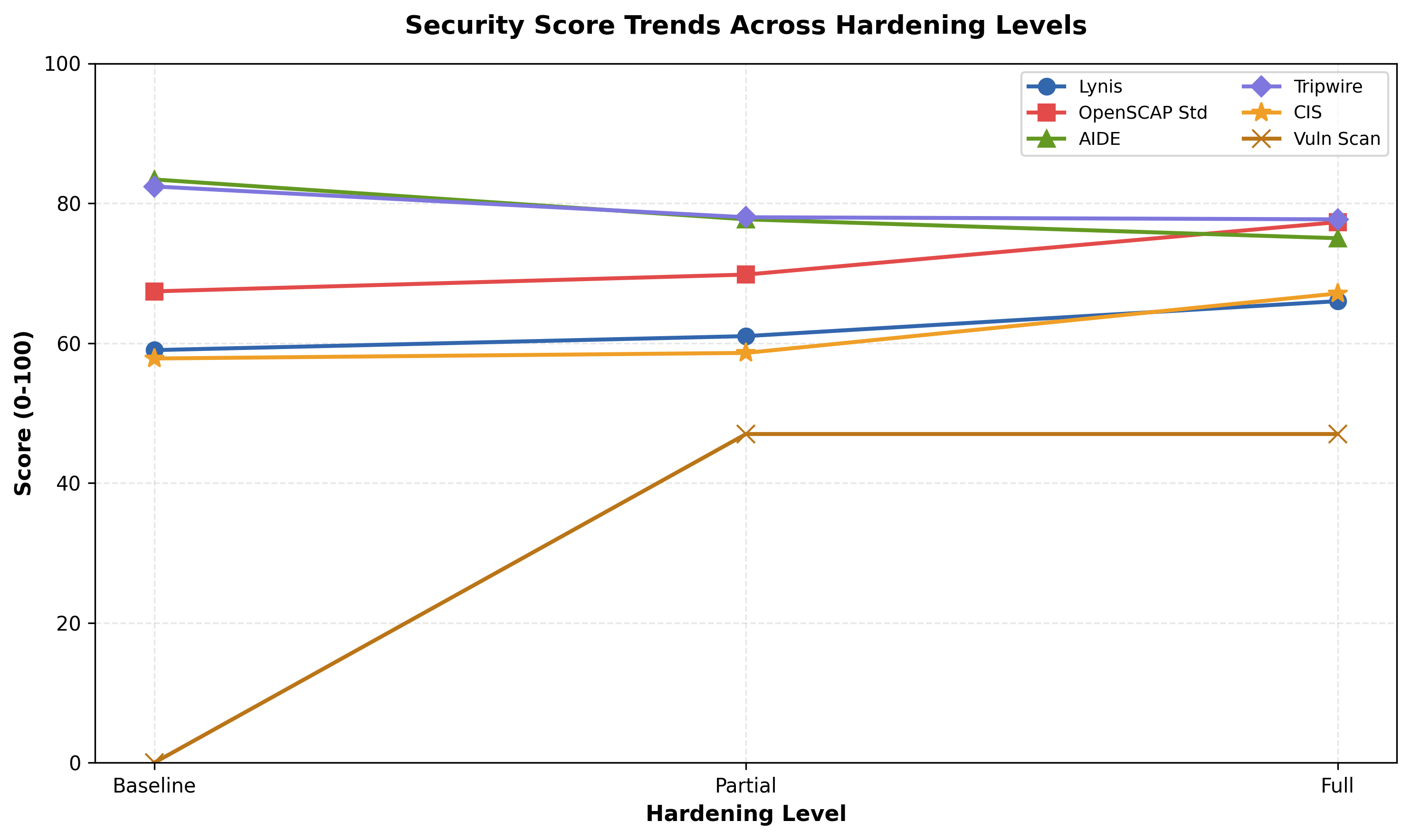}
\caption{Score trends revealing three patterns: compliance
tools trending upward, file integrity tools trending
downward, and vulnerability scanning showing a step-function
improvement at firewall activation.}
\label{fig:trends}
\end{figure}

The divergent tool behavior validates multi-tool aggregation.
An organization using only Lynis concludes hardening is
working; using only AIDE suggests instability. Only
aggregation reveals the complete picture: genuine security
improvement with expected filesystem changes as a
consequence.

\subsection{Cross-Platform Validation}

\begin{table}[h]
\centering
\caption{Cross-Platform Score Comparison
(Baseline / Partial / Full)}
\label{tab:crossval}
\begin{tabular}{lll}
\toprule
\textbf{Tool} & \textbf{Local VM} & \textbf{FABRIC} \\
\midrule
Lynis & 61 / 63 / 68 & 59 / 61 / 66 \\
OpenSCAP Standard & 76.7 / 69.7 / 77.2 &
67.4 / 69.8 / 77.3 \\
OpenSCAP CIS & 64.3 / 60.7 / 68.3 &
57.8 / 58.6 / 67.1 \\
Tripwire & 98.7 / 93.3 / 98.3 &
82.4 / 78.0 / 77.7 \\
\bottomrule
\end{tabular}
\end{table}

Absolute scores differ due to platform-specific
configurations but both environments exhibit consistent
improvement trends, confirming platform-independent
methodology.

\section{Case Studies}

\subsection{Case Study 1: Basic Apache Web Server}

Apache 2.4.52 was installed on the baseline node (10.10.1.1)
with default configuration. Hardening included: ServerTokens
Prod, ServerSignature Off, TRACE disabled, security headers
(X-Content-Type-Options, X-Frame-Options, X-XSS-Protection,
CSP, Referrer-Policy, HSTS), directory listing disabled,
autoindex/status modules removed, UFW for SSH/HTTP/HTTPS.
UCA improved from 49.36 to 53.53 (+8.4\%).

\subsection{Case Study 2: DVWA Web Application Server}

\subsubsection{Target Selection}

We selected DVWA to balance reproducibility and system-level access requirements. DVWA \cite{dvwa}, endorsed
by OWASP with 8,000+ GitHub stars and citations in hundreds
of academic papers, was selected. Metasploitable~2/3 were
rejected because their legacy OS versions (Ubuntu 8.04/14.04)
lack compatible SCAP content. External platforms were rejected
because only nmap can scan them remotely.

\subsubsection{DVWA Vulnerability Profile}

DVWA provides documented vulnerabilities configurable at four
security levels. This study used level ``low'' (maximally
vulnerable):

\begin{itemize}[noitemsep]
\item SQL Injection and Blind SQL Injection
\item XSS: reflected, stored, DOM-based
\item Command Injection; File Inclusion (LFI/RFI)
\item Unrestricted File Upload; CSRF; Brute Force
\end{itemize}

\subsubsection{Measurement Scope}

The UCA framework measures OS-level and network-level
security posture. DVWA's PHP application vulnerabilities
(SQLi, XSS, command injection) are invisible to Lynis,
OpenSCAP, AIDE, and Tripwire, which assess the Ubuntu 22.04
host configuration. The nmap scan operates at the network
layer, detecting open ports, service CVEs, and
misconfigurations, including an exposed \texttt{.git}
directory confirmed at baseline. This case study evaluates
the server hosting DVWA, not the PHP application itself,
consistent with the framework's design scope.

\subsubsection{Setup}

DVWA was deployed on node 10.10.1.4 (Ubuntu 22.04, 4~cores,
8~GB RAM) with Apache~2.4.52, PHP~8.1.2, and
MariaDB~10.6.23, cloned from the official GitHub repository.
All six UCA tools were installed. AIDE and Tripwire databases
were initialized once before any hardening.

\subsubsection{Hardening Applied}

\textbf{SSH:} Root login/password auth disabled,
MaxAuthTries~3, X11/TCP forwarding disabled.
\textbf{Firewall:} UFW deny incoming, allow 22/80/443;
reduced open ports from 3 to 2.
\textbf{Kernel:} Full sysctl hardening (IP forwarding off,
SYN cookies, ASLR max, SUID dump off).
\textbf{System:} auditd with web-specific rules; password
aging; filesystem blacklist; core dumps disabled.
\textbf{Apache:} ServerTokens Prod, ServerSignature Off,
TraceEnable Off, security headers, autoindex/status disabled.
\textbf{PHP:} expose\_php Off, display\_errors Off.
\textbf{Remediation:} Exposed \texttt{.git} directory
removed.

\subsubsection{Results and Analysis}

\begin{table}[H]
\centering
\caption{Case Study Results Comparison}
\label{tab:case}
\begin{tabular}{lcccccc}
\toprule
& \multicolumn{2}{c}{\textbf{Basic Web Server}}
& & \multicolumn{2}{c}{\textbf{DVWA Server}} \\
\cmidrule{2-3} \cmidrule{5-6}
\textbf{Tool} & \textbf{Before} & \textbf{After}
& & \textbf{Before} & \textbf{After} \\
\midrule
Lynis & 58 & 65 & & 60 & 67 \\
OpenSCAP Std & 67.4 & 77.3 & & 60.5 & 70.5 \\
AIDE & 52.6 & 52.6 & & 84.2 & 73.8 \\
Tripwire & 55.2 & 51.9 & & 88.6 & 72.0 \\
OpenSCAP CIS & 57.4 & 66.3 & & 57.0 & 66.3 \\
Vulnerability & 0 & 0 & & 81 & 94 \\
\midrule
\textbf{UCA} & \textbf{49.36} & \textbf{53.53}
& & \textbf{70.55} & \textbf{73.20} \\
\textbf{Gain} & \multicolumn{2}{c}{\textbf{+8.4\%}}
& & \multicolumn{2}{c}{\textbf{+3.8\%}} \\
\bottomrule
\end{tabular}
\end{table}

\begin{figure}[H]
\centering
\includegraphics[width=\textwidth]{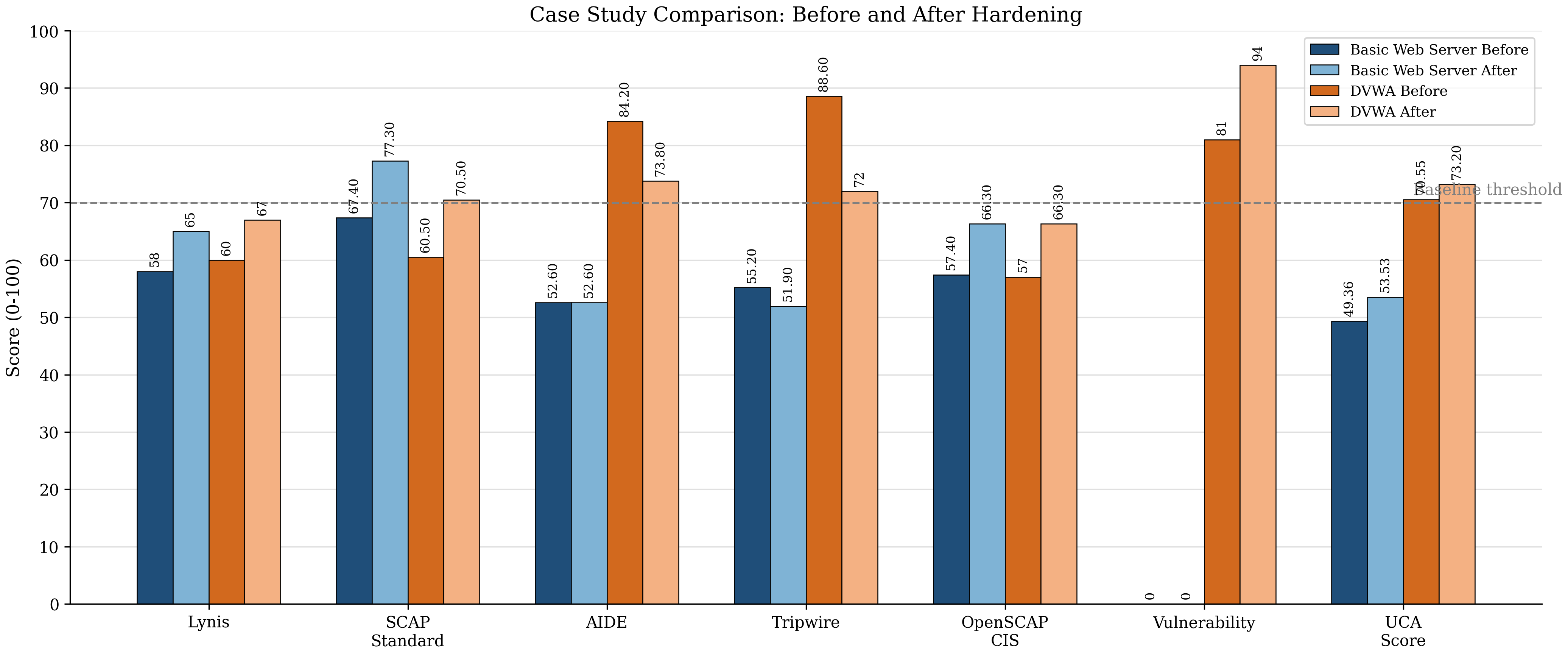}
\caption{Grouped bar comparison of individual tool scores and
composite UCA scores before and after hardening for both case
studies. The DVWA server starts at a higher baseline (70.55 vs
49.36) reflecting Ubuntu~22.04 defaults with a full LAMP
stack. Compliance tools improve in both deployments. File
integrity tools decline from hardening-induced changes. The
vulnerability score improves by 13 points on the DVWA server
through firewall activation and \texttt{.git} removal. UCA
composite improves +8.4\% (basic) and +3.8\% (DVWA),
consistent with diminishing returns at higher baselines.}
\label{fig:case}
\end{figure}

\textbf{Compliance improvement:} Lynis +7, OpenSCAP Standard
+10.0, CIS +9.3 confirm hardening addressed requirements
across multiple frameworks simultaneously.

\textbf{File integrity decline:} AIDE $-$10.4 and Tripwire
$-$16.6 correctly detected hardening-induced filesystem
changes. This is expected and consistent with Experiment~1.

\textbf{Vulnerability improvement:} Score rose from 81 to 94
(+13 pts) through firewall activation and \texttt{.git}
removal. Remaining CVEs require package upgrades beyond
configuration scope.

\textbf{Diminishing returns:} The DVWA node started at 70.55
because Ubuntu~22.04 with a LAMP stack satisfies many
compliance rules by default. The 3.8\% net improvement
reflects diminishing returns at higher security levels.
Individual tool improvements of up to +13 points confirm the
framework detects real security gains.

\section{Discussion}

\subsection{Framework Effectiveness}

Across all three experiments the UCA framework detected
improvements from 3.8\% to 16.8\% with individual tool
changes spanning $-$16.6 to +47 points. Weighted aggregation
correctly balanced opposing trends, producing net positive
composites reflecting genuine improvement. The multi-tool
approach generated insights unavailable to single-tool
assessment: firewall as the dominant driver (71.7\%), file
integrity cross-validation, and quantified diminishing
returns across deployment types.

\subsection{Comparison with Previous Work}

\begin{table}[h]
\centering
\caption{Evolution from UCA v1.0 to v2.0}
\label{tab:evolution}
\begin{tabular}{lll}
\toprule
\textbf{Feature} & \textbf{v1.0 \cite{paul2025arxiv}}
& \textbf{v2.0} \\
\midrule
Tools & 3 & 6 \\
Weights & 0.4/0.4/0.2 & 0.15-0.20 \\
Platform & FABRIC EDUKY & FABRIC FIU + VMware \\
File Integrity & AIDE only & AIDE + Tripwire \\
Compliance & Standard only & Standard + CIS \\
Vulnerability & None & nmap NSE \\
AIDE Scoring & Linear & Logarithmic \\
UCA Improvement & +25.9\% & +16.8\% \\
Case Studies & None & 2 (web server, DVWA) \\
\bottomrule
\end{tabular}
\end{table}

The larger v1.0 improvement reflects a lower EDUKY baseline
(39.73\% vs 67.4\% OpenSCAP) and custom rule scoring. V2.0
exchanges statistical depth for broader coverage,
vulnerability assessment, cross-platform validation, and
practical case studies.

\subsection{Principal findings}

\begin{enumerate}[noitemsep]
\item Progressive hardening produced +16.8\% UCA improvement
(58.34 to 68.17), with compliance tools gaining 11.9-16.1\%
and file integrity tools correctly detecting changes.

\item Firewall activation contributed +47 vulnerability
points (71.7\% of total), invisible without vulnerability
scanning integration.

\item The logarithmic AIDE formula prevented score collapse
across 317 cumulative changes, resolving v1.0's primary
limitation.

\item Cross-platform validation on VMware and FABRIC confirmed
platform-independent methodology.

\item The DVWA case study achieved +3.8\% net improvement with
individual tool gains up to +13 points, demonstrating
applicability to web application server environments and
confirming diminishing returns at higher baselines.

\item Divergent tool behavior provides empirical validation to suggest 
that no single tool captures the complete security picture.
\end{enumerate}

All tools, configurations, and methodologies are open-source
and documented for reproduction and extension.

\subsection{Limitations and Validity Assessment}

\textbf{OpenVAS unavailability:} GVM~21.4 CLI errors required
nmap substitution, reducing CVSS granularity.
\textbf{Static weights:} Manual assignment; ML-based
optimization planned.
\textbf{Version CVEs:} OpenSSH~8.9p1 and Apache~2.4.52 CVEs
persist post-hardening; require package upgrades.
\textbf{Single distribution:} Ubuntu~22.04 only; SCAP content
requires adaptation for other distributions.
\textbf{Statistical depth:} Single scans per configuration;
multiple runs would strengthen claims.
\textbf{OS vs application layer:} DVWA's PHP vulnerabilities
are outside UCA scope; application security tool integration
would extend coverage.

One important limitation in this paper that we acknowledge and can significantly impact the validity of the results is that the experiments were reported after one cycle. Their is a need to execute those experiments several times in different environments to ensure stability of results. 

\section{Future Work}

\textbf{LLM Integration:} Microsoft Phi-3 and Meta Llama~3
can analyze UCA outputs, generate risk explanations, and
prioritize remediation by predicted score impact.

\textbf{Adaptive Weighting:} Random Forest classifier trained
on experimental data to optimize weights per deployment
context.

\textbf{Automated Remediation:} Scripts generating prioritized
remediation commands from UCA output.

\textbf{Statistical Validation:} Multiple independent runs
with t-tests and effect size calculations.

\textbf{Cross-Distribution and Cloud:} CentOS, Debian, Rocky
Linux; AWS, Azure, GCP.

\textbf{Continuous Monitoring:} Scheduled scans with alerting
below defined thresholds.

\textbf{Application-Layer Integration:} Dedicated web
application security tools extending UCA to OWASP Top~10
scope.

\section{Conclusion}

This paper presented a unified framework for aggregating outputs from multiple security tools into a single composite score. The framework improves interpretability of security assessments and provides a practical approach for evaluating system hardening.

While further validation is needed, the results suggest that multi-tool aggregation can offer useful insights into system security posture.

\bibliographystyle{plain}

\begin{thebibliography}{21}

\bibitem{nist800-53}
NIST, ``Security and Privacy Controls for Information Systems
and Organizations,'' SP~800-53 Rev.~5, 2020.

\bibitem{ibm2023}
IBM Security, ``Cost of a Data Breach Report 2023,'' IBM
Corporation, 2023.

\bibitem{verizon2023}
Verizon, ``2023 Data Breach Investigations Report,'' Verizon
Enterprise Solutions, 2023.

\bibitem{paul2025arxiv}
S. Paul and I. Alsmadi, ``Security Hardening Using FABRIC:
Implementing a Unified Compliance Aggregator,''
arXiv:2601.00909, 2025.

\bibitem{fabric2019}
I. Baldin et al., ``FABRIC: A National-Scale Programmable
Experimental Network Infrastructure,'' \textit{IEEE Internet
Computing}, vol.~23, no.~6, pp.~38-47, 2019.

\bibitem{lynis}
CISOfy, ``Lynis Security Auditing Tool,''
\url{https://cisofy.com/lynis/}

\bibitem{scap}
D. Waltermire et al., ``Technical Specification for SCAP,''
NIST SP~800-126 Rev.~3, 2018.

\bibitem{openscap}
OpenSCAP Project, ``OpenSCAP NIST Certified Toolkit,''
\url{https://www.open-scap.org/}

\bibitem{ssg}
ComplianceAsCode, ``SCAP Security Guide,''
\url{https://github.com/ComplianceAsCode/content}

\bibitem{aide}
R. Lehti and P. Virolainen, ``AIDE,''
\url{https://aide.github.io/}

\bibitem{tripwire}
Tripwire Inc., ``Open Source Tripwire,''
\url{https://github.com/Tripwire/tripwire-open-source}

\bibitem{cis}
Center for Internet Security, ``CIS Benchmarks,''
\url{https://www.cisecurity.org/cis-benchmarks}

\bibitem{nmap}
G. Lyon, \textit{Nmap Network Scanning}, Nmap Project, 2009.

\bibitem{openvas}
Greenbone Networks, ``OpenVAS,''
\url{https://www.openvas.org/}

\bibitem{owasp}
OWASP Foundation, ``OWASP Top Ten,''
\url{https://owasp.org/Top10/}

\bibitem{dvwa}
DVWA Project, ``Damn Vulnerable Web Application,''
\url{https://github.com/digininja/DVWA}

\bibitem{siem}
A. Chuvakin, K. Schmidt, and C. Phillips, \textit{Logging
and Log Management}, Syngress, 2012.

\bibitem{soar}
Gartner, ``Market Guide for SOAR Solutions,'' 2019.

\bibitem{multicriteria}
A. Shameli-Sendi et al., ``Taxonomy of ISRA,''
\textit{Computers \& Security}, vol.~57, pp.~14-30, 2016.

\bibitem{autocomp}
M. Schwartz et al., ``Automated Security Compliance
Assessment,'' \textit{Bell Labs Technical Journal}, vol.~12,
no.~3, pp.~203-218, 2007.

\bibitem{commercial_comparison}
A. Alhomoud et al., ``A Survey on Vulnerability Assessment
Tools,'' \textit{ICICS}, pp.~1--6, 2011.

\end{thebibliography}

\end{document}